\documentclass[twoside,fleqn]{article}
\usepackage{espcrc2}
\usepackage{graphicx}

\title{
{\Large  QCD Correlators at Large $N_C$}
}
\author{J.J. Sanz-Cillero  
\address{
       {\em ,} Groupe Physique Th\'eorique, IPN Orsay \\
       {\em Universit\'e Paris-Sud XI,  91406 Orsay, France}}
       } 
       
\begin{document}

\newcommand{\xs}{\mbox{$x(\sigma)$}}
\newcommand{\vs}{\vbox{\vskip 1cm plus .3cm minus .3cm}}
\def\question#1{{{\marginpar{\small \sc #1}}}}
\newcommand{\bean}{\begin{eqnarray*}}
\newcommand{\eean}{\end{eqnarray*}}
\newcommand{\gapproxeq}{\lower
.7ex\hbox{$\;\stackrel{\textstyle >}{\sim}\;$}}
\newcommand{\lapproxeq}{\lower
.7ex\hbox{$\;\stackrel{\textstyle <}{\sim}\;$}}
\newcommand{\arr}{\mbox{$\rightarrow$} }
\newcommand{\Apr}{\mbox{\overline{ \rm p} } }
\newcommand\lsim{\mathrel{\rlap{\lower4pt\hbox{\hskip1pt$\sim$}}
    \raise1pt\hbox{$<$}}}
\newcommand\gsim{\mathrel{\rlap{\lower4pt\hbox{\hskip1pt$\sim$}}
    \raise1pt\hbox{$>$}}}
\newcommand{\ba}{\begin{array}}
\newcommand{\ea}{\end{array}}
\newcommand{\nn}{\nonumber}
\newcommand{\mathbold}{\bf}
\newcommand{\be}{\begin{equation}}
\newcommand{\ee}{\end{equation}}
\newcommand{\bear}{\begin{eqnarray}}
\newcommand{\eear}{\end{eqnarray}}
\newcommand{\tab}{\hspace*{0.5cm}}
\newcommand{\cen}{\hspace*{7.0cm}}
\newcommand{\ii}{\'{\i}}
\newcommand{\II}{\'{\I}}
\newcommand{\sla}{\hspace*{-0.2cm}\slash  }
\newcommand{\slag}{\hspace*{-0.25cm} \slash}

\newcommand{\rvac}{\,|0\rangle}
\newcommand{\lvac}{\langle 0|\,}
\newcommand{\ket}{\,\rangle}
\newcommand{\bra}{\langle \,}
\newcommand{\eqn}[1]{(\ref{#1})}
\newcommand{\cO}{{\cal O}}
\newcommand{\bel}[1]{\be\label{#1}}
\newcommand{\mL}{\mathcal{L}}
\newcommand{\mA}{\mathcal{A}}
\newcommand{\mB}{\mathcal{B}}
\newcommand{\mC}{\mathcal{C}}
\newcommand{\mD}{\mathcal{D}}
\newcommand{\mM}{\mathcal{M}}
\newcommand{\mO}{\mathcal{O}}
\newcommand{\mF}{\mathcal{F}}
\newcommand{\mT}{\mathcal{T}}
\newcommand{\mR}{\mathcal{R}}
\newcommand{\mG}{\mathcal{G}}
\newcommand{\Frac}[2]{\frac{\displaystyle #1}{\displaystyle #2}}
\newcommand{\Int}{\displaystyle{\int}}

\begin{abstract}
This talk explores the spin--1  correlators  up to 
$\cO(\alpha_s)$ through a large $N_C$ resonance theory.  
The phenomenological analyses of this kind  
must take these corrections into account since they produce a larger
impact than  the first OPE
condensates.  It is also  necessary to separate  low and high energy regimes; 
fixing  the parameters of the lightest multiplets  
through perturbative QCD arguments is unfair and introduces errors in the
determination of the condensates and resonance parameters. 
This separation of regimes improves our 
 understanding of the Minimal Hadronical Approximation.  
The study  at  $\cO(\alpha_s)$
already allows   discerning  between  
different  hadronical models, where the Regge-like mass spectrum 
shows the best agreement to phenomenology. 
\end{abstract}

\maketitle




%
%
%
\section{Introduction}
\tab This talk explores the spin--1 QCD correlators\footnote{
Talk given at QCD'05, 
July 4-9th  2005, Montpellier (France). 
Further details can be found in Ref.~\cite{correlator}.  
 This work was partially supported by    
EU~RTN~Contract~CT2002-0311.
}.  
The Operator Product Expansion (OPE)  has 
resulted a very powerful and successful 
instrument to describe Quantum Chromodynamics (QCD) in the  
domain of deep euclidean  
momenta~\cite{Shifman}; in addition to   
the purely perturbative QCD contributions (pQCD) one finds non-trivial 
operators of higher dimension. Likewise, 
in the large $N_C$ limit 
\, \,  --being $N_C$ the number of colours--, QCD suffers 
large simplifications~\cite{NC}   
(this limit of QCD   will be denoted as $QCD_\infty$). 
Assuming  confinement,    
$QCD_\infty$ is dual to a theory  with an infinite
number of hadronic states where  
the processes are  given by the tree-level topologies. 
The theory  in terms of mesons will be denoted as 
Resonance Chiral Theory ($R\chi T$)~\cite{therole}.  
It must be built up  chiral 
invariant in order to ensure the right low energy 
dynamics~\cite{chpt} 
even at the loop 
level~\cite{quantumloops}.

The   correlators are defined as
\be
\ba{l} 
(q_\mu q_\nu\, - \, q^2 g_{\mu\nu}) \, \Pi_{_{XY}}(q^2)\, = \, 
\\ 
\tab \tab \tab i\, \Int dx^4 \, e^{iqx} \, \bra T\{J_{_{X}}(x)_\mu \, 
J_{_{Y}}(0)_\nu^\dagger \}\ket 
\, ,
\ea
\ee  
denoting $J_X^\mu$ and $J_Y^\nu$   
either vector or axial-vector current.   
We actually analyse the V-A  
and V+A combinations 
($\Pi_{_{LR}}=\Pi_{_{VV}}-\Pi_{_{AA}}$ and 
$\Pi_{_{LL}}=\Pi_{_{VV}}+\Pi_{_{AA}}$). 
Only the  sector of light quarks $u/d/s$ will be considered and 
within the chiral and large $N_C$ limits.

Several  large $N_C$ resonance models have studied the $V+A$ correlator 
at the free-quark level~\cite{LVS,toymodel,5D}.  
The analysis is taken here with detail up to $\cO(\alpha_s)$, profiting the
relation between pQCD, the resonance couplings and the mass 
spectrum~\cite{correlator,EspriuRegge}. One  needs then 
to introduce a separation scale 
(here taken as $M_p^2\sim $~2~GeV$^2$)  
to split perturbative and non-perturbative dynamics~\cite{correlator}. 

Some models for the resonance
mass spectrum are explored,  
getting a set of predictions for the
$\rho(770)$ and $a_1(1260)$ parameters and the first OPE condensates. 
The Minimal Hadronical Approximation of $R\chi T$ 
(MHA)~\cite{therole,KPdR} arises 
as a  well justified approach to $QCD_\infty$ that, however, introduces 
truncation errors that must be properly estimated.

\section{pQCD and OPE}
\tab 
At ${Q^2\equiv -q^2 \gg
\Lambda_{_{QCD}}^2}$ the  $V-A$ correlator 
is well described by the OPE~\cite{Shifman}:
\be 
\Pi_{_{LR}}^{^{OPE}}\,\,\,=
\,\,\,  \,\,\, \sum_{m=3}^\infty\Frac{\bra \cO_{_{(2m)}}^{^{LR}}\ket}{Q^{2m}} 
\,  .  
\ee
where the coefficients   
$\bra \cO_{(2m)}\ket$ are provided by  the dimension--$(2m)$ operator 
in the OPE.  The   $V+A$    combination   shows the structure~\cite{Shifman} 
\bel{eq.PILLRGE}
\Pi_{_{LL}}^{^{OPE}} \,\, \, \, = \Pi_{_{LL}}^{^{pQCD}} 
\,\,\, \, + \,\,\, \sum_{m=2}^\infty 
\Frac{\bra \cO_{(2m)}^{^{LL}}\ket}{Q^{2m}}\, ,
\ee
with the OPE series starting by  
the pQCD contribution    $\Pi_{_{LL}}^{^{pQCD}}$.

\section{Resonance theory at large $N_C$}
\tab
At large $N_C$, the correlators are  given by an infinite exchange of
narrow-width resonances,
\bel{eq.RChT1}
\ba{rl}
\Pi_{_{LR}}^{^{N_C\to\infty}}=& 
-\Frac{F^2}{Q^2}+ 
\displaystyle{\sum_{j=1}^{\infty}} 
\Frac{[-\pi_j]\, F_{j}^2}{M_{j}^2+Q^2} \, ,
\\
\Pi_{_{LL}}^{^{N_C\to\infty}}=& 
\,\, 
\Frac{F^2}{Q^2}+ 
\displaystyle{\sum_{j=1}^{\infty}} 
\Frac{F_{j}^2}{M_{j}^2+Q^2} \, ,  
\ea
\ee 
where $M_j$ and $F_j$ are the mass and coupling constant of the $j$--th 
spin--1 multiplet  at LO in $1/N_C$, and $F$  is 
the pion decay constant. 
The factor $\pi_j$ denotes the parity of the corresponding multiplet. 
We will consider  an alternating parity spectrum ($\pi_j=(-1)^{j}$) 
and  $M_1\leq M_2\leq...$

\subsection{Separation of perturbative and non-perturbative QCD contributions}
\tab 
First of  all,   one has    
to impose a separation scale, taken in this work as  $M_p^2\sim 2$~GeV$^2$,    
between the perturbative and non-perturbative regimes:
\be 
{\Pi^{^{N_C\to\infty}}=\Delta\Pi^{^{N_C\to\infty}}_{_{pert.}}
+\Delta\Pi^{^{N_C\to\infty}}_{_{non-p.}}} \, .
\ee 
where the perturbative sub-series 
$\Delta\Pi_{_{pert.}}^{^{N_C\to\infty}}$  
includes the resonances with $M_{j}^2 > M_p^2$, and   
the non-perturbative part   
$\Delta\Pi_{_{non-p.}}^{^{N_C\to\infty}}$ 
contains  those with   
$M_{j}^2 \leq M_p^2$.

It is also necessary  
to make some assumption about  the
mass spectrum at high energies. A smooth $M_j^2$
dependence on $j$ will be assumed for $M_j^2>M_p^2$. 
If the couplings $F_j$ also follow a smooth
behaviour at high energies  then, in order to recover $\Pi_{_{LL}}^{^{pQCD}}$ 
at $Q^2\gg \Lambda_{_{QCD}}^2$, they must 
behave like~\cite{correlator,EspriuRegge} 
\bel{eq.FjIMPI}
F_{j}^2  =   \delta M_j^2 \cdot 
\left[ \Frac{1}{\pi}\mbox{Im}\Pi_{_{LL}}(M_j^2)^{^{pQCD}}  
+ \cO\left(\Frac{1}{M_j^2}\right)\right] \, , 
\ee
being $\delta M_j^2\equiv M_j^2-M_{j-1}^2$. 
The $\cO\left(\frac{1}{M_j^2}\right)$ corrections were neglected in this work. 
Hence, at ${Q^2\gg M_p^2}$,   
the perturbative sub-series results  
\bel{eq.OPEpert}
\ba{rl}
\Delta\Pi_{_{pert.}}^{^{N_C\to\infty}}
=& 
\Pi^{^{pQCD}}+  \displaystyle{\sum_{m=1}^\infty} \Frac{\Delta\bra
\cO_{_{(2m)}}\ket^{^{pert.}}}{Q^{2m}}\, , 
\ea
\ee
with $\Pi_{_{LR}}^{^{pQCD}}=0$.  
Remark that the  $\Delta\bra \cO_{_{(2m)}}^{^{LL}}\ket^{^{pert.}}$ 
cannot be recovered in general  by trivially expanding 
Eq.~\eqn{eq.RChT1} in powers of $\frac{M_R^2}{Q^2}$.

On the other hand, the non-perturbative part of the series produces just 
$1/Q^{2m}$ power terms at $Q^2\gg M_p^2$, yielding the corresponding 
contributions to the condensates, e.g.  
${\Delta\bra \cO_{_{(2m)}}^{^{LL}}\ket^{^{non-p.}}=(-1)^{m-1} 
\,\sum_{j=1}^p  F_j^2 \,(M_j^2)^{m-1} \, .}
$

Matching the sum 
$\bra \cO_{_{(2m)}}\ket
 = \Delta \bra \cO_{_{(2m)}}\ket^{^{pert.}} 
 + \Delta \bra \cO_{_{(2m)}}\ket^{^{non-p.}}  $  
to the OPE one gets the familiar 
\bel{eq.constraint1} 
\bra \cO_{_{(2)}}^{^{LL}}\ket =\bra \cO_{_{(2)}}^{^{LR}}\ket=
\bra \cO_{_{(4)}}^{^{LR}}\ket=0 \, .  
\ee

\section{Mass spectrum  models: phenomenology} 
\tab
Several models for the mass spectrum  are  
studied in Ref.~\cite{correlator} with detail. 
The multiplets with $M_n^2\gsim$~2~GeV$^2$ ($n\geq 3$) 
are included in $\Delta \Pi^{^{N_C\to\infty}}_{_{pert.}}$. 
An asymptotic $M_n^2$
dependence is taken for them and 
their masses are set such that  
${M_3^2=M_{\rho'}^2\simeq (1.45}$~GeV$)^2$ and 
${M_4^2=M_{a_1'}^2\simeq (1.64}$~GeV$)^2$~\cite{PDG}.

Given the asymptotic mass spectrum the couplings are fixed through
Eq.~\eqn{eq.FjIMPI} and  the combination     
$\left(\Delta \Pi^{^{N_C\to\infty}}_{_{pert.}}-\Pi^{^{pQCD}}\right)$
shows the structure 
$\displaystyle{\sum_{m=1}^\infty} 
\frac{\Delta\bra\cO_{_{(2m)}}\ket^{^{pert.}}}{Q^{2m}}$. 
The corresponding  
$\Delta\bra\cO_{_{(2m)}}\ket^{^{pert.}}$   are recovered  
through a numerical analysis for the range  
$Q^2=2-6$~GeV$^2$~\cite{correlator}.       
pQCD is only taken up to $\cO(\alpha_s)$  so 
$\cO(\alpha_s^2)$ uncertainties are expected in the derivation of the 
condensates.  
 
The pion, $\rho(770)$ and $a_1(1260)$  are included into   
$\Delta \Pi^{^{N_C\to\infty}}_{_{non-p.}}$. 
We take  $F=92.4$~MeV and  $M_\rho=0.77$~GeV as inputs. 
The remaining parameters,      
$F_\rho,\, F_{a_1}$ and $M_{a_1}$,  are recovered     
through the matching to OPE in Eq.~\eqn{eq.constraint1}. 
This fixes completely the correlators, 
producing a prediction for    
$\bra\cO_{_{(4)}}^{^{LL}}\ket$ and 
$\bra\cO_{_{(6)}}^{^{LR}}\ket$. 
The results  for the two most relevant models are shown  
in Table~\eqn{tab.predictions}  and the corresponding 
$\Pi_{_{LR}}$ and   
$\mA_{_{LL}}=-Q^2\Frac{d\Pi_{_{LL}}}{dQ^2}$  
are compared in Fig.~\eqn{fig.adlerRT} to OPE~\cite{narisonpid}.

\subsection{\bf $QCD_\infty$ in $1+1$ dimensions}
\tab 
In several models, like $QCD_{\infty}$  in $1+1$ dimensions 
and Regge theory,   
the high energy spectrum follows the dependence  
$M_n^2=\Lambda^2 +n \,\delta M^2$~\cite{NC,LVS,toymodel,EspriuRegge}.   
The analysis yields  $\rho $ and $a_1$ couplings 
of the same order as the asymptotic values $F_\rho\sim
F_{a_1}\sim F_{j\geq 3}$. Both resonance couplings 
and the $a_1$ mass are in agreement with former   
analyses~\cite{therole,PDG,spin1fields,exprespar}.

\begin{table}
\hspace*{-0.5cm}
\begin{tabular}{|c|c|c|}
\hline
\rule[-0.7em]{0em}{1.9em}
& 1+1 $QCD_\infty$ &
5D--spectrum
\\
 \hline\rule{0em}{1.2em}
$F_\rho$ (MeV) & 
$139\pm 3$  & 
 $174^{+5}_{-7}$    
\\
 \hline\rule{0em}{1.2em}
$F_{a_1}$ (MeV) &  
$138\pm 3 $ & 
$169^{+6}_{-8} $   
\\
 \hline\rule{0em}{1.2em}
$M_{a_1}$ (MeV) & 
$1180^{+17}_{-19}$ & 
 $1029^{+19}_{-13}$     
\\
 \hline\rule{0em}{1.2em}
$\bra\cO_{_{(4)}}^{^{LL}}\ket$   ($10^{-3}$~GeV$^4$)& 
$-30^{+22}_{-30}$ & 
$100^{+70}_{-60}$    
\\
 \hline\rule{0em}{1.2em}
$\bra\cO_{_{(6)}}^{^{LR}}\ket$   ($10^{-3}$~GeV$^6$)
& 
$-3.7^{+0.8}_{-0.3}$ & 
$0.3^{+2.2}_{-1.5}$   
\\
\hline\end{tabular}
\caption{\small Predictions for the two first resonance multiplets  
within each mass spectrum model. } 
\vspace*{-1em}
\label{tab.predictions}
\end{table}
\begin{figure}[t!]
\begin{center}
\includegraphics[width=6cm,clip]{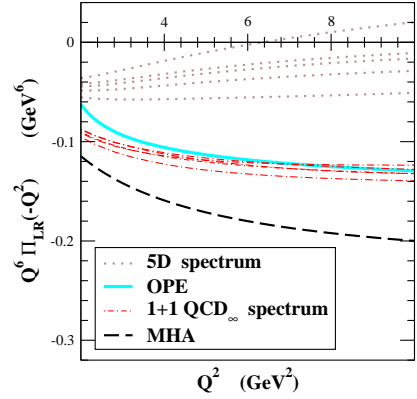}
\includegraphics[width=6cm,clip]{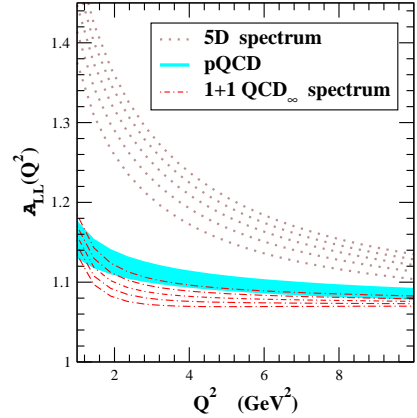}
\caption{\small 
$V-A$ correlator and $V+A$ Adler function 
in R$\chi T$ up to 
$\cO(\alpha_s)$  and within MHA 
($M_{a_1}=M_\rho/\sqrt{2}, \, F_\rho/\sqrt{2}=F_{a_1}=F$~\cite{spin1fields}). 
They are compared to the results from the OPE and pQCD at that
order~\cite{narisonpid}.}
\label{fig.adlerRT}
\end{center}
\end{figure}

\subsection{Five--dimensional spectrum}
\tab 
Another available scenario is provided by models 
in five dimensions~\cite{5D}, which produce 
a four dimensional  effective spectrum with dependence 
$M_n=\Lambda+ n \, \delta M$.  
One finds an acceptable value for $M_{a_1}$ although   
$F_\rho$ and $F_{a_1}$ go high above the usual
determinations~\cite{therole,PDG,spin1fields,exprespar}. 
Other sum-rule analyses show that this spectrum 
produces a clearly distinctive pattern which differs from 
the Regge-like model and from the one found in the
phenomenology~\cite{correlator}.

\subsection{Minimal Hadronical Approximation}
\tab 
The perturbative contributions to the condensates 
are found in general of the order of magnitude of the usual hadronical 
parameters --the pion decay constant and the light resonance masses--.    
For instance, for the Regge-like spectrum one gets  
\bel{eq.reg}
\ba{l}
\Delta \bra\cO_{_{(2m)}}^{^{LL}}\ket^{^{pert}} 
\,\simeq \Frac{(-1)^{m}}{m} \, 
\Frac{N_C}{12\pi^2} \, \left(M_{\rho'}^2\right)^m
\, , 
\\
\Delta\bra\cO_{_{(2m)}}^{^{LR}}\ket^{^{pert}}\,
 \simeq  (-1)^{m-1} 
\Frac{N_C}{24\pi^2} \delta M^2 
 \left(M_{\rho'}^2\right)^{m-1}
  \, ,  
\ea
\ee
up to corrections suppressed by $\frac{\delta M^2}{M_{\rho'}^2}$.
Actually, from Eq.~\eqn{eq.reg} one finds for the Regge spectrum 
that at euclidean momenta the perturbative sub-series 
$\Delta \Pi_{_{LR,\, pert.}}^{^{n_C\to\infty}}$ is 
approximately equivalent to the contribution from  
an effective  $\rho'$ resonance 
with coupling ${\widetilde{F}_{\rho'}^2\equiv \Frac{N_C}{24\pi^2} \delta
M^2}$, resulting of the order of   $F^2$.
This exemplifies how some spectrum patterns provide a description equivalent
to that from a theory with a finite number of multiplets.

In the MHA~\cite{KPdR}, 
$\Delta \Pi_{_{pert.}}^{^{N_C\to\infty}}$ is taken as zero, introducing a 
truncation error $\cO\left(\delta M^2 \, (M_{\rho'}^2)^{m-1}\right)$  
in the value of the condensates. The 
short distance matching for  $\Pi_{_{LR}}$ under MHA  is then 
perfectly justified as long as  one is 
aware of the inherent uncertainties: 
\bel{eq.WSR}
\ba{rl}
F^2 \,- \, F_\rho^2\,+ \, F_{a_1}^2\, &=\,
\cO\left(\frac{N_C}{24\pi^2}\delta M^2\right) \, , 
\\ 
F_{a_1}^2 \, M_{a_1}^2  
\, - \, F_\rho^2\, M_\rho^2\, 
\, &=  \cO\left(\frac{N_C}{24\pi^2}\delta M^2\, M_{\rho'}^2\right)\, .
\ea
\ee

The resonance theory  with a finite number of multiplets  
allows to build a proper  
quantum field theory for resonances with a well defined $1/N_C$ perturbative
expansion and renormalisation scheme, being 
renormalisable order by order in $1/N_C$~\cite{quantumloops}. 
The smooth variation of the resonance parameters from MHA to
MHA+$\rho'$~\cite{SRFriot}  
seems to hint the close relation between the truncated $R\chi T$ 
and the full  $QCD_\infty$.

\section{Conclusions}
\tab 
It was shown 
how $R\chi T$ is able to recover 
the OPE up to $\cO(\alpha_s)$.
The analysis up to this order  is 
essential for any $V+A$ phenomenological analysis since the pQCD 
corrections are more important than those from the first OPE 
condensates.

In order to handle 
the infinite tower of hadronic states 
one needs  to assume a structure $M_j^2$ 
for the spectrum at high energies and 
that the $F_j^2$ depend smoothly on $j$. 
The matching to $\Pi_{_{LL}}^{^{pQCD}}$ in the deep euclidean imposes then  
strong constraints on the asymptotic structure of the couplings. 
some of the current models are tested. 
1+1~$QCD_\infty$ ($M_n^2\sim n$) 
and  the 5D theories ($M_n^2\sim n^2$)  
show the closest  agreement to phenomenology, 
providing the first one the best description. Nevertheless, estimates on
subleading corrections are still necessary.

The scale $M_p^2\sim 2$~GeV$^2$ splits  the perturbative and
the non-perturbative regimes,  so the pion, $\rho$ and $a_1$ are included in the
non-perturbative sub-series.    
This splitting allows understanding how 
$R\chi T$ under MHA    
connects the full $QCD_\infty$.  
Dropping the contribution   
from $\Delta \Pi_{_{pert.}}^{^{NC\to\infty}}$ 
is a well defined approximation as long as 
the truncation uncertainties are properly considered. 

{\large \bf Acknowledgments:  }
I would like  to thank S.~Narison 
for the organisation of QCD'05. 

%
%
%
%
%
%
%
%
%
%

\end{document}